\documentclass{appolb}
\usepackage{graphicx}

\usepackage[super,compress]{cite}
\usepackage{graphicx}
\usepackage{xspace}
\usepackage{comment}
\usepackage{url}
\usepackage{hyperref}
\hypersetup{colorlinks,citecolor=black,filecolor=black,linkcolor=black,urlcolor=blue}

\usepackage{amsmath}
\usepackage{comment}

\newcommand{\dd}{ {\textrm d}}
\newcommand{\ee}{ {\textrm e}}

\begin{document}
\title{Speed of sound in Kaluza\,--\,Klein Fermi gas%
}
\author{Anna Horv\'ath,$^{1,2}$, Emese Forgács-Dajka,$^{2,3}$, \\ Gergely G\'abor Barnaf\"oldi$^{1}$
\address{$^{1}$HUN-REN Wigner RCP, 29-33 Konkoly-Thege M. Str., 1121 Budapest, Hungary}
\\
\address{$^{2}$E\"otv\"os Lor\'and University, P\'azm\'any P\'eter stny. 1/A, 1117 Budapest, Hungary}
\\
\address{$^{3}$HUN-REN-SZTE Stellar Astrophysics Research Group,  766 Szegedi Rd., 6500 Baja, Hungary}
}

\maketitle
\begin{abstract}
A five-dimensional Kaluza-Klein spacetime model is considered, with one extra compactified spatial dimension. The equation of state of an electrically neutral, zero-temperature Fermi gas with a repulsive linear potential is described. From the equation of state, the speed of sound squared is calculated and shown for different model parameters. Its properties are studied from lower energies up to the conformal limit. 
\end{abstract}
  
\section{Introduction}

One of the biggest challenges today is to determine the equation of state of extreme, superdense matter present in compact astrophysical objects or at the early stage of the Universe. Modeling the extreme matter results in a huge number of equation-of-state variants~\cite{Altiparmak:2022bke}. Parameters of these theoretical models can be set via fitting to astronomical observation or high-energy collision data~\cite{Ozel:2016oaf,PhysRevLett.121.161101,Gardim:2019xjs,Tsang:2018kqj,Wei_2020}. Typically the theoretical model parameters with physical meaning cannot be directly determined in a phenomenological way. But some of them is accessible to these tests, like the speed of sound squared ($c^2_\mathrm{s}$), is which connected to the slope of the equation of state~\cite{Marczenko:2024uit,Altiparmak:2022bke,PhysRevC.107.025802}.

Investigating matter at extreme conditions provides the possibility of probing new unified theories or physics beyond the Standard Model (BSM). This motivated us to test compact stars in Kaluza\,--\,Klein spacetime with a compactified extra dimension~\cite{Karsai:2016wfx, Barnafoldi:2015wca,Barnafoldi:2007, Barnafoldi:2020phq, Horvath_2024}. Within this model a simplified equation of state was used to determine the size of the compactification radius of an extra dimension --- constrained by pulsar data. 
Following the idea presented by Horv\'ath {\it et al}~\cite{Horvath_2024}, the sensitivity of the speed-of-sound squared on model parameters is studied here with interesting aspects.

\section{The Kaluza\,--\,Klein Spacetime and the Equation of State}
\label{sec:KK}

The Kaluza\,--\,Klein theory originates in the first part of the 20\textsuperscript{th} century, when Theodor Kaluza extended spacetime with one extra spatial dimension, to unify gravity with electromagnetism in a geometrical way~\cite{Kaluza:1921tu}. Later, Oskar Klein extended the model by assuming that the additional dimension is compactified~\cite{Klein:1926tv}. In this model the spacetime has $R^4\times S^1$ topology: the usual 1 time-like and 3 infinite spatial dimensions are complemented with a finite-sized spatial one, which is curled up in a circle, introducing a periodic boundary condition. Since the size of this extra dimension is taken to be microscopic (on the order of a few fm in our case), it is naturally explained why its existence is non-detectable in everyday circumstances --- too high of an energy would be needed. Since the advent of the theory, it has been studied in various works and served as a base to many other models, such as string theory~\cite{Overduin:1997sri,Veneziano:1968yb}. 

In the simplest case~\cite{Barnafoldi:2007,Horvath_2024}, one can consider a diagonal five-dimensional metric tensor,
\begin{equation}
    g_{AB} = \text{diag}(g_{00},g_{11},g_{22},g_{33},g_{55}) \ ,
\end{equation}
where the metric functions $g_{00},...,g_{33}$ correspond to the usual 4-dimensional metric describing gravity, while $g_{55}$ denotes the fifth dimension\footnote{Note, traditionally the 4\textsuperscript{th} spatial coordinate is denoted by index 5 for being of the fifth compactified dimension.}, and is often associated with a scalar field~\cite{Coquereaux:1990qs}. In this work, we follow Ref.~\cite{Horvath_2024}, we take the fifth-dimensional metric component to be a constant, thus the effects of the modified spacetime can be compressed into a mass spectrum of particles moving in the extra dimension with energy 
\begin{equation}
    E = \sqrt{\mathbf{k}^2+k_\mathrm{5}^2+m^2} = 
    \sqrt{\mathbf{k}^2+\bar{m}^2}\ .
    \label{eq:energy}
\end{equation}
Here $\mathbf{k}$ is the usual 3-dimensional, while $k_{\mathrm{5}}$ is the 5-dimensional momentum, $m$ is the particle's mass and $c=\hbar=1$. Thus, from a 4-dimensional point of view, particles possess a modified, effective mass, $\bar{m}^2(N_\mathrm{exc}) = m^2 + k_\mathrm{5}^2$, where $ k_5= N_\mathrm{exc}/{r_\mathrm{c}}$. Here, the size of the extra dimension is denoted by $r_\mathrm{c}$, and the energy level corresponding to the five-dimensional movement is represented by the excitation number, $N_\mathrm{exc}$.  

To describe the thermodynamical properties of a medium, where particles can have Kaluza\,--\,Klein excitations, the equation of state of an zero-temperature, non-interacting Fermi gas was used. Particles need to have high-enough energies in order to propagate in the microscopic compactified dimension. In cold nuclear matter, which can be found in neutron stars at extreme densities~\cite{Ozel:2016oaf}, the zero-temperature limit is a good approximation~\cite{Glendenning:1997wn}. In these density regimes, the effects of the strong force become important at short ranges, thus we introduced a simple, repulsive potential, 
\begin{equation}
    U(n) = \xi n \ ,
    \label{eq:poti}
\end{equation}
which is linear in the baryon density, $n$, to model the "nuclear-like" interaction between baryons. The strength of the 5-dimensional interaction, $\xi$ can be fixed from nuclear physics arguments~\cite{Zimanyi:1987bt}, however, here it remains a model parameter. After deriving the equations of state from the thermodynamic potential, the interaction potential appears in the state variables through a modified chemical potential, $\bar{\mu}=\mu-U(n)$, and as an interaction term in the pressure and the energy density: $p_\mathrm{int}=\varepsilon_\mathrm{int}=\int U(n)dn = \frac{1}{2}\xi n^2$. A more detailed description about the equation of state can be read eg. in Horváth {\it et. al.}~\cite{Horvath_2024}, where graphs and comparison to other models are also shown.

\section{Results on the speed of sound}
\label{sec:cs2}

The equation of state is usually provided by the function-form of the thermodynamic state variables. The energy-density, $\varepsilon(p(\mu))$ or the pressure $p(\varepsilon(\mu))$ are directly connected to the speed of sound by, 
\begin{equation}
    c_\mathrm{s}^2=\left. \frac{\partial p}{\partial \varepsilon} \right| _{s} \ , 
\end{equation}
which is sensitive to the different model parameters: the size of the extra dimension, $r_\mathrm{c}$, the interaction strength, $\xi$ and the excitation number, $N_\mathrm{exc}$. Latter denotes the highest energy level a particle was allowed to occupy. As a ground-state baryon for the spectrum, the neutron ($n^0$) with a mass of $m=939.57$~MeV was used and the masses of baryons with higher excitations, $\bar{m}$, follow the so called Kaluza\,--\,Klein ladder. 
\begin{figure}[!ht]
    \centering
    \includegraphics[scale=0.6]{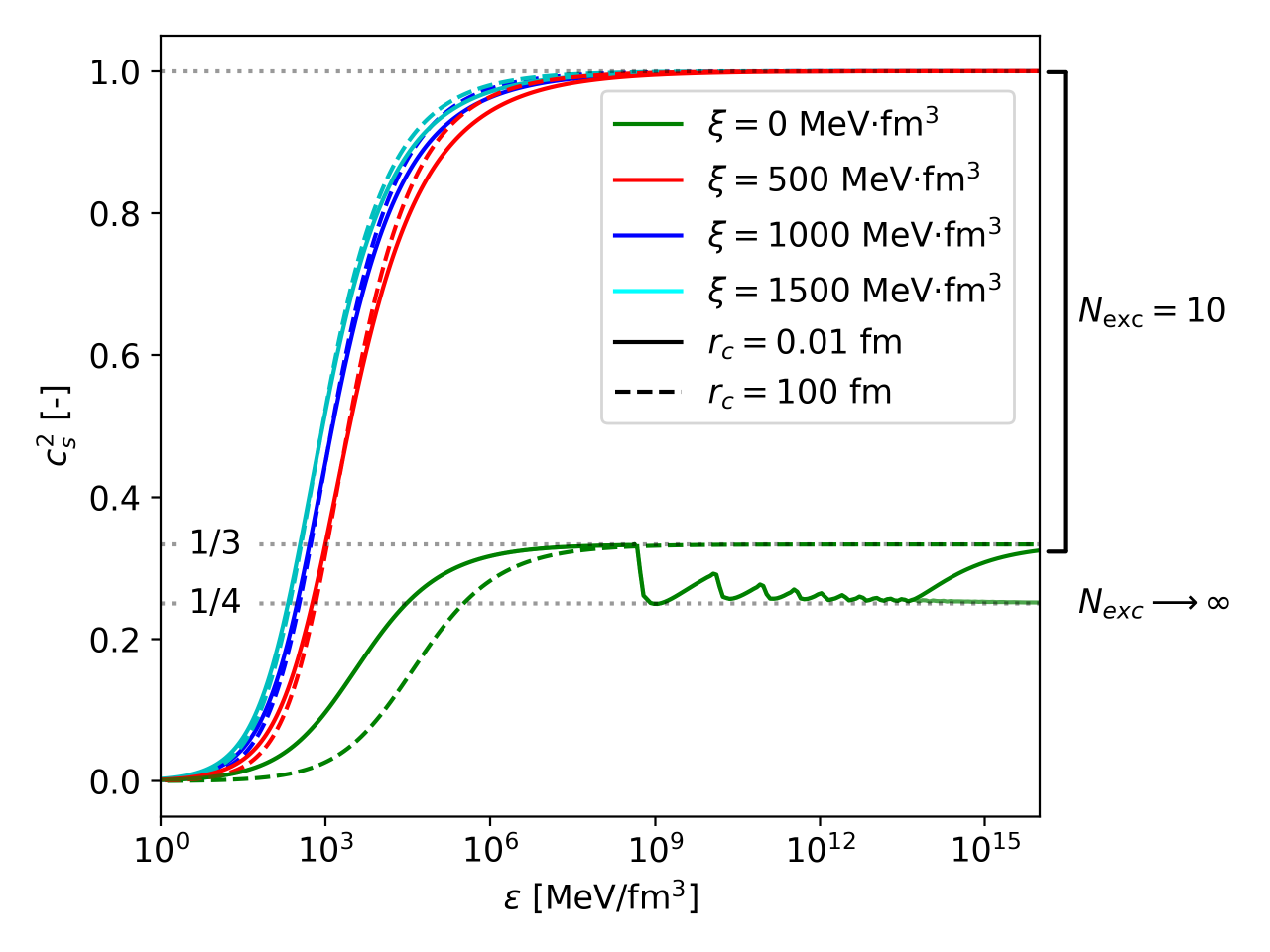}
    \caption{Speed of sound squared as a function of energy density, $c_\mathrm{s}^2 (\varepsilon)$. (Colors: different interaction strength, $\xi$. Line style: size of the extra dimension, $r_\mathrm{c}$.)}
    \label{fig:cs}
\end{figure}

Figure~\ref{fig:cs} shows the speed of sound squared as a function of energy density. Different colors correspond to different interaction strengths, $\xi$. One can see that the non-interacting case is qualitatively distinct from the interacting ones at higher energies. In general, the speed-of-sound squared starts from zero and increases until saturation, reached at around $10^6-10^8$~MeV/fm\textsuperscript{3}, which energy regime is where the BSM region starts. However, we note that in a realistic neutron star model the relevant energy scales do not reach the $10^4$~MeV/fm\textsuperscript{3} regime. In our recent study~\cite{Horvath_2024}, specifically, the central energy-density goes up to about $\varepsilon_\mathrm{c}=1100\pm150$~MeV/fm\textsuperscript{3} for a realistic maximal mass, stable pulsar having about two solar masses. 

The interacting cases reach unity for all the interaction strength values: $\xi=1500$~MeV/fm\textsuperscript{3}, $\xi=1000$~MeV/fm\textsuperscript{3} and $\xi=500$~MeV/fm\textsuperscript{3} shown by cyan, blue and red curves respectively. The non-interacting cases (green lines) approach lower conformal limit values. 

Considering the size-variation of the extra compactified dimension, the smaller $\xi$, the more prominent the effect of the extra dimension -- the curves corresponding to $r_\mathrm{c}=0.01$~fm (solid) and $100$~fm (dashed) differ more significantly. The effect of $r_\mathrm{c}$ is most visible if one sets the interaction strength to $\xi=0$, which manifests in a two orders of magnitude difference between the saturation energy of the two $r_\mathrm{c}$ cases -- higher for $r_\mathrm{c}=100$~fm.

Calculations were performed for two cases with respect to the number of possible excited states. We evaluated the speed of sound at a finite $N_\textrm{exc}$ value, and for a case with infinite degrees of freedom. 
\begin{description}
    \item[$N_\textrm{exc}=10$:] 
    For $\xi=0$ and $r_\mathrm{c}=0.01$~fm, 10 dips are clearly visible, which correspond to the appearance of the new degrees of freedom, new energy levels. Since the number of excited states is fixed at $N_\textrm{exc}=10$, after reaching the tenth level, the effect of the extra dimension slowly disappears, and curves tend to the 3-dimensional case, where the conformal limit, $c_\mathrm{s}^2=1/3$ is approached. 
    \item[$N_\textrm{exc}=\infty$:] 
    The asymptotic behavior is modified at the extreme high energy limit, if we allow an infinite number of energy levels to open up. Due to a phase-space extended by the extra dimension, the conformal limit becomes $c_\mathrm{s}^2=1/4$, as shown in light green. 
\end{description}
One can also see that until the energy of the first excited state is reached, the theory is simply 3-dimensional. Dips are not present for the other cases, because either a large $\xi$ is dominating them, or -- for $r_\mathrm{c}=100$~fm -- the extra dimensional spectrum essentially becomes a continuum, thus not exhibiting discrete behavior.

The significant difference between the interacting and the non-interacting case can be explained by the simplicity of the repulsive potential. At large energies (higher densities) it "blows up", resulting in a very stiff equation of state. Applying a slightly more complex potential would show a more realistic behavior, however, at energies relevant for neutron stars, the equation of state is still quite well-behaved. Causality holds for all cases: the speed of sound does not exceed the speed of light.   

\section{Summary}

In this work, the speed of sound described by a 5-dimensional, zero-temperature Fermi-gas equation of state is studied, relevant for neutron stars. First, we showed how the theory is built up, then performed numerical calculations for analysis. The equation of state respects causality, and --- in the non-interacting case --- it shows how the conformal limit changes for a higher-dimensional theory. Significant dips caused by the availability of new energy levels in the extra dimension are also present, depending on the size of the extra dimension and the strength of the repulsive potential. 

\section*{Acknowledgment}
Authors gratefully acknowledges the Hungarian National Research, Development and Innovation Office (NKFIH) under Contracts No. OTKA K147131, No. NKFIH 2024-1.2.5-TÉT-2024-00022 and Wigner Scientific Computing Laboratory (WSCLAB). Author A. H. is supported by the DKÖP program of the Doctoral School of Physics of Eötvös Loránd University and EU Erasmus fellowship. E.F-D. also received funding from the NKFIH excellence grant TKP2021-NKTA-64 and HUN-REN.

\bibliographystyle{IEEEtran}

\bibliography{multidimkk}


\end{document}